\documentclass[twocolumn,showpacs,preprintnumbers,amsmath,amssymb,superscriptaddress,prb]{revtex4}
\usepackage{graphicx}% Include figure files
\usepackage{dcolumn}% Align table columns on decimal point
\usepackage{bm}% bold math
\begin{document}

\newcommand{\avg}[1]{\langle #1 \rangle}
\newcommand{\var}[1]{\text{Var} (#1)}
\newcommand{\cov}{\text{Cov}}

\title{Effect of Electric Field on Diffusion in Disordered Materials\\
 I. One-dimensional Hopping Transport}

\author {A. V. Nenashev}
%\email {nenashev@isp.nsc.ru}
\affiliation{Institute of Semiconductor Physics, 630090 Novosibirsk, Russia}
\affiliation{Novosibirsk State University, 630090 Novosibirsk, Russia}

\author {F. Jansson}
\email {fjansson@abo.fi}
\affiliation{Graduate School of Materials Research,
\AA bo Akademi University,  20500 Turku, Finland}
\affiliation{Department of Physics and Center for Functional Materials,
\AA bo Akademi University, 20500 Turku, Finland}

\author {S. D. Baranovskii}
\affiliation{Department of Physics and Material Sciences Center,
Philipps-University, 35032 Marburg, Germany}

\author {R. \"Osterbacka}
\affiliation{Department of Physics and Center for Functional Materials,
\AA bo Akademi University, 20500 Turku, Finland}

\author {A. V. Dvurechenskii}
\affiliation{Institute of Semiconductor Physics, 630090 Novosibirsk, Russia}
\affiliation{Novosibirsk State University, 630090 Novosibirsk, Russia}

\author{F. Gebhard}
\affiliation{Department of Physics and Material Sciences Center,
Philipps-University, 35032 Marburg, Germany}

\date{\today}

\begin{abstract}
An exact analytical theory is developed for calculating the diffusion
coefficient of charge carriers in strongly anisotropic disordered
solids with one-dimensional hopping transport mode for any
dependence of the hopping rates on space and energy. So far such
a theory existed only for calculating the carrier mobility.
The dependence of the diffusion
coefficient on the electric field evidences a linear, non-analytic
behavior at low fields for all considered models of disorder.
The mobility, on the contrary, demonstrates a parabolic, analytic
field dependence for a random-barrier model, being linear,
non-analytic for a random energy model. For both models the
Einstein relation between the diffusion coefficient and mobility
is proven to be violated at any finite electric field. The
question on whether these non-analytic field dependences of the
transport coefficients and the concomitant violation of the
Einstein's formula are due to the dimensionality of space or due
to the considered models of disorder is resolved in the following
paper [Nenashev et al., arXiv:0912.3169], where analytical
calculations and computer simulations are carried out for two- and
three-dimensional systems.
\end{abstract}

\pacs{72.20.Ht, 72.20.Ee, 72.80.Ng, 72.80.Le}
\keywords{hopping transport, mobility, diffusion, Einstein relation,
multiple trapping, Gaussian disorder model}

%Use showkeys class option if keyword
%72.20.Ht High-field and nonlinear effects
%72.20.Ee Mobility edges; hopping transport
%72.80.Ng Disordered solids
%72.80.Le Polymers; organic compounds (including organic semiconductors)

\maketitle

\section{Introduction}
\label {sec-introduction} Charge carrier transport in disordered
materials - inorganic, organic and biological systems - has been
in the focus of intensive experimental and theoretical study for
several decades due to various current and potential applications
of such materials in modern electronic devices (see, for instance,
Ref.~\onlinecite{Baranovski2006} and references therein).  An
essential part of the research is dedicated to studying the
mobility of the charge carriers, $\mu$, and their diffusion
coefficient, $D$, as the decisive transport coefficients
responsible for the performance of most devices. Among other features,
the relation between these two transport coefficients is the
subject of intensive research, since this relation (called the
``Einstein relation'') often provides significant information on
the underlying transport mechanism.\cite{Baranovski2006} In
numerous experimental studies on organic disordered materials,
essential deviations from the conventional form
\begin {equation}
\label {eq-Einstein}
\mu = \frac{e}{kT} D,
\end {equation}
of this relation have been recognized.
\cite{Yuh1988,Borsenberger1991,Borsenberger1993b, Hirao1996,
Hirao1997, Lupton2002, Harada2005} In Eq.~(\ref{eq-Einstein}) $e$
is the elementary charge, $T$ is temperature and $k$ is the
Boltzmann constant.  Einstein derived this relation between
$\mu$ and $D$ for the case of thermal equilibrium in a
non-degenerate system of charge carriers. Deviations from
Eq.~(\ref{eq-Einstein}) were predicted theoretically for
non-equilibrium transport at low temperatures
\cite{Baranovskii1996,Baranovskii1998,Baranovskii1998b} and also
for equilibrium transport in degenerate systems if the density of
states (DOS), which can be used by the charge carriers, strongly
depends on energy, for instance, exponentially \cite{Ritter1988}
or according to a Gaussian distribution.\cite{Roichman2002}
Usually the former DOS is assumed for inorganic amorphous
semiconductors, while the latter one is assumed for disordered
organic materials, such as molecularly doped and conjugated
polymers.\cite{Roth1991,Bassler2000,Pope1999,Bassler1993,%
Borsenberger1993,Auweraer1994}
In this paper and in the following one (Ref. \onlinecite{Nenashev2009II})
we derive general equations for calculating the
diffusion coefficient and the mobility of charge carriers and
apply them to systems with the Gaussian DOS, since most
of the experimental evidence for the violation of Eq.~(\ref{eq-Einstein}) has
been reported for organic disordered materials. The DOS is taken
as
\begin {equation}
\label {eq-gaussian-dos} g (\varepsilon) = \frac {N} {\sigma \sqrt{2
\pi}} \exp \left( - \frac {\varepsilon^2} {2 \sigma^2} \right),
\end {equation}
where $N$ is the spatial concentration of conducting states and
$\sigma$ is the energy scale of the DOS distribution.

Remarkably, experiments on disordered organic
materials evidence that at relatively low electric fields, at
which the carrier mobility $\mu$ is field-independent and hence
the carrier transport can be treated as Ohmic one (low-field
regime), the diffusion coefficient $D$ of charge carriers and
concomitantly the relation between $\mu$ and $D$ become
essentially dependent on the magnitude of the applied electric
field $F$.\cite{Yuh1988,Borsenberger1993b,Hirao1996,Hirao1997} Our
aim in this paper and in the following one\cite{Nenashev2009II} 
is to provide an analytical theory  for the
field-dependent diffusion coefficient and mobility of charge
carriers. The theory will be checked by computer simulations.

Charge transport in disordered organic materials is dominated by
incoherent hopping of electrons and holes via localized states
randomly distributed in space, with the DOS described by
Eq.~(\ref{eq-gaussian-dos}).\cite{Roth1991,Bassler2000,Pope1999,%
Bassler1993,Borsenberger1993,Auweraer1994}
The transition rate between an occupied state $i$ and an empty state
$j$, separated by the distance $r_{ij}$, is described by the
Miller-Abrahams expression \cite{Miller1960}
\begin {equation}
\label{eq-Miller-Abrahams}
\Gamma_{ij} = \nu_0 \, e^{-2\frac{r_{ij}}{a}} \left\{
\begin{array}{ll}
 e^{ - \frac{\Delta \varepsilon_{ij}}{kT}} 
\qquad & ,\ \Delta \varepsilon_{ij} > 0\\
 1  & ,\ \Delta \varepsilon_{ij} \leq 0
  \end{array}  \right. ,
\end {equation}
where $\nu_0$ is the attempt-to-escape frequency.
The energy difference between the sites is
\begin {equation}
\label{eq-delta} \Delta \varepsilon_{ij} = \varepsilon_j - \varepsilon_i -
F e (x_j - x_i),
\end {equation}
where the electric field $F$ is assumed to be directed
along the $x$-direction.  The localization length of the charge
carriers in the states contributing to hopping transport is $a$. We
assume the latter quantity to be independent of energy and we will
neglect correlations between the energies of
the localized states, following the Gaussian-disorder-model
of B\"assler. \cite {Bassler1993,Borsenberger1993,Auweraer1994}

The challenging problem arises of how to describe theoretically
the field-dependent diffusion of charge carriers in the hopping regime
within the Gaussian DOS.  This very problem was addressed in the
numerical simulations by Richert \emph{et al.}\cite{Richert1989} Using a
Monte Carlo algorithm with a randomly distributed parameter $a$
(the so-called off-diagonal disorder), it was shown that the diffusion
coefficient for hopping transport in the Gaussian DOS depends
essentially on the field strength at such low electric fields that the
mobility of charge carriers remains
field-independent.\cite{Richert1989} This result was interpreted in
analytical calculations by Bouchaud and Georges,\cite{Bouchaud1989}
who considered a hopping process in a one-dimensional (1D) system of
equidistant localized states with transition rates essentially
different from those given by Eq.~(\ref{eq-Miller-Abrahams}). In the
calculations of Bouchaud and Georges \cite{Bouchaud1989} the
transition rates between the neighboring sites were taken as
\begin{equation}
\label{eq-rate-RBM}
\Gamma_{i, i\pm1} = \Gamma_0 
\exp \left[ \frac {\Delta_{i\pm1,i} \pm eFd}{2kT} \right]
\end{equation}
with $\Delta_{i,i+1} = \Delta_{i+1,i}$ distributed according to
$g(\Delta_{ij})$ given by Eq.~(\ref{eq-gaussian-dos}). We will call
this model the Random-Barrier-Model (RBM) in contrast to the model
described by Eqs.~(\ref{eq-gaussian-dos}) and
(\ref{eq-Miller-Abrahams}), which we call the Random-Energy-Model
(REM).  Bouchaud and Georges \cite{Bouchaud1989} suggested for the
field-dependent part of the diffusion coefficient in the RBM the
expression $D(F) - D(0) \propto F \exp[3 \sigma^2 / 8(kT)^2]$, which
they claimed to be precisely the dependence found in
Ref.~\onlinecite{Richert1989}. Later the authors of
Ref.~\onlinecite{Richert1989} studied the quantity $D(F) - D(0)$ by
computer simulations in more detail \cite{Pautmeier1991} and found a
quadratic dependence of $D(F) - D(0)$ on $F$ at low fields and no
turn-over to a linear field dependence as suggested by Bouchaud and
Georges.\cite{Bouchaud1989} The question arises then on whether this
discrepancy in the field dependences of the diffusion constant between
the computer simulations \cite{Pautmeier1991} and analytical calculations
\cite{Bouchaud1989} is due to different models (RBM
\cite{Bouchaud1989} against REM \cite{Pautmeier1991}), or it is due to
different dimensionalities considered in these two approaches (1D in
analytical calculations \cite{Bouchaud1989} against 3D in computer
simulations \cite{Pautmeier1991}).  The only way to answer this
question is to obtain exact results for the REM in 1D and to compare
them with the results for the RBM in 1D on one hand and with the
results for the REM in 3D on the other hand. This task demands
developing a new analytical method for calculating drift and diffusion
in 1D systems for the hopping transport mode.  In
Sec.~\ref{sec-method} we present such method. We also present in
Sec.~\ref{sec-analytic-RBM} the exact result for the
field-dependent diffusion in the RBM, which differs from the one given
by Bouchaud and Georges.\cite{Bouchaud1989}
Sec.~\ref{sec-analytic-REM} is devoted to analytical results on the
field-dependent diffusion coefficient and mobility in the REM in the
1D case. The exact results for both RBM and REM give a linear field
dependence of the diffusion coefficient at low fields. In
Sec.~\ref{sec-numeric} we present the results obtained by computer
simulations in 1D systems. Concluding remarks are gathered in
Sec.~\ref{sec-conclusions}.

The following paper\cite{Nenashev2009II}
is devoted to diffusion in 3D systems. 
The results in the 3D case clearly
demonstrate a quadratic field dependence of the diffusion
coefficient at low fields.
One should then conclude that the
discrepancy between the linear \cite{Bouchaud1989} and the
quadratic \cite{Pautmeier1991} field dependences of the diffusion
constant reported in the literature is due to the different space
dimensionalities considered in the two approaches. One should note
that the differences between 1D systems and 3D systems with
respect to the field-dependent diffusion coefficient have been
reported in the literature, albeit for systems with essential
correlations between energies and spatial positions of localized
states involved into the hopping transport. Relying essentially on
such correlations, Parris \emph{et al.} \cite{Parris1997} obtained an
exact result for the field-dependent diffusion coefficient in 1D
systems, which was not confirmed in computer simulations carried
out on 3D correlated systems. \cite{Novikov2006} Our study leads
to a similar conclusion for the Gaussian disorder model without
space-energy correlations.
This study was necessary, since the theory from Ref.~\onlinecite{Parris1997}
cannot be applied to the case of uncorrelated disorder.

\section{Analytical method}
\label{sec-method} This section is devoted to one-dimensional
hopping in the presence of an electric field. The considered system
consists of a chain of sites separated by a constant distance $d$.
Each site is either empty or occupied by a carrier. We consider
the limit of small carrier concentration, therefore the
probability for the $i$th site to be occupied, $p_i$, is small for
each $i$. The time evolution of probabilities $p_i$ is described
by equation
\begin{equation} \label{eq:2-time-evolution}
\frac{\partial p_i}{\partial t} = \Gamma_{i-1,i}\, p_{i-1} +
\Gamma_{i+1,i}\, p_{i+1} - (\Gamma_{i,i-1}+\Gamma_{i,i+1})\, p_i,
\end{equation}
where $\Gamma_{ij}$ is the rate of transition from site $i$ to site $j$.
Transition rates $\Gamma_{ij}$ are assumed to be time-independent; to
be non-zero only for nearest neighbors $(\Gamma_{ij}\neq0
\Leftrightarrow |i-j|=1)$; and to obey the principle of detailed
balance:
\begin{equation} \label{eq:2-detailed-balance}
\frac{\Gamma_{i,i+1}}{\Gamma_{i+1,i}} = \exp \frac{\varepsilon_i-\varepsilon_{i+1}+eFd}{kT},
\end{equation}
where $\varepsilon_i$ is the energy of a carrier on the $i$th site
\emph{without} the electric field, and $F$ is the electric field strength.

Our aim is to obtain analytical expressions for diffusion
coefficients with transition rates $\Gamma_{ij}$ chosen according
to either RBM or REM. A similar problem was considered by
Derrida\cite{Derrida1983} who obtained exact results for diffusion
coefficient in \emph{finite} systems with arbitrarily chosen
transition rates. But, in the limit of an \emph{infinite} system,
his expression (Eq.~(47) of Ref.~\onlinecite{Derrida1983})
contains an uncertainty of type ``$\infty-\infty$'', and resolving
this uncertainty is a non-trivial task. Derrida considered an
infinite system only for the case if $\Gamma_{ij}$ are random
\emph{independent} variables, except that only $\Gamma_{ij}$ and
$\Gamma_{ji}$ may be correlated. This condition is fulfilled for
the RBM, but not for the REM, in which $\Gamma_{ij}$ and
$\Gamma_{jk}$ are correlated due to the common site $j$. 
Therefore Derrida's approach can hardly be generalized
to for the REM.
Here we propose another analytical approach for
evaluating the diffusion coefficient in the infinite disordered
one-dimensional systems. Derrida's method uses a definition of the
diffusion coefficient $D$ related to random walks:
\begin{equation} \label{eq:2-def-D1}
 D = \frac12 \lim_{t\rightarrow\infty}\frac{d}{dt}\left( \langle x^2(t)\rangle
 - \langle x(t)\rangle^2 \right),
\end{equation}
where $x(t)$ is the position of the particle at time $t$. On the
contrary, our method is based on the macroscopic definition of $D$
as a ratio of current flow $j$ and the long-scale gradient of the
concentration $n$ of particles:
\begin{equation} \label{eq:2-def-D2}
 D = - \frac{j(x)}{dn(x)/dx}.
\end{equation}
We believe that both methods give the same results, though our
method has an advantage of providing an \emph{explicit} expression
for $D$ in the general case of the \emph{infinite} one-dimensional
system (see
Eqs.~(\ref{eq:2-vD-from-ab}), (\ref{eq:2-solution-for-a}), and
(\ref{eq:2-solution-for-b})
below). This expression can be straightforwardly applied to the
particular cases of the RBM and REM.

We start by considering the continuous-medium approximation. This
approximation deals with the carrier concentration $n(x,t)$
averaged upon a sufficiently large spatial scale. The time
evolution of this concentration obeys the Fokker-Planck equation
\begin{equation} \label{eq:2-Fokker-Planck}
\frac{\partial n}{\partial t} = - v \frac{\partial n}{\partial x}
 + D \frac{\partial^2 n}{\partial x^2},
\end{equation}
provided that $n$ varies in space sufficiently slowly (that is,
the characteristic scale of spatial variation is large as compared
to the scale of averaging). Here $v$ is the drift velocity and $D$
is the diffusion coefficient. Let us consider the initial
concentration $n(x,0)$ in the form
\begin{equation} \label{eq:2-exp-mu-x}
n(x,0) = n_0 \exp(\eta x)
\end{equation}
with an infinitely small factor $\eta$. The solution of
Eq.~(\ref{eq:2-Fokker-Planck}) with the initial
condition~(\ref{eq:2-exp-mu-x}) reads
\begin{equation} \label{eq:2-exp-mu-x-lambda-t}
n(x,t) = n_0 \exp(\eta x - \lambda t),
\end{equation}
where
\begin{equation} \label{eq:2-lambda-from-mu}
\lambda = v \eta - D \eta^2.
\end{equation}
Since both $\lambda$ and $\eta$ are infinitely small, one can resolve
Eq.~(\ref{eq:2-lambda-from-mu}) with respect to $\eta$ in the following
way:
\begin{equation} \label{eq:2-mu-from-lambda}
\eta = \frac1v \lambda + \frac{D}{v^3} \lambda^2 + O(\lambda^3).
\end{equation}
We will use Eq.~(\ref{eq:2-mu-from-lambda}) for calculating the
drift velocity $v$ and the diffusion coefficient $D$. For this
aim, we need a \emph{microscopic} definition of the
coefficients $\lambda$ and $\eta$ expressed in terms of occupation
probabilities $p_i$ rather than in terms of the concentration $n$.

To obtain an exponential time dependence of the concentration,
$n\sim\exp(-\lambda t)$, we can simply postulate that each probability
$p_i$ depends on time in the same way, $p_i\sim\exp(-\lambda t)$.
Therefore $\partial p_i / \partial t = -\lambda p_i$, and
Eq.~(\ref{eq:2-time-evolution}) can be written as
\begin{equation} \label{eq:2-time-evolution-lambda}
-\lambda p_i = \Gamma_{i-1,i} p_{i-1} + \Gamma_{i+1,i} p_{i+1}
- (\Gamma_{i,i-1}+\Gamma_{i,i+1})\, p_i.
\end{equation}
This is the way of introducing $\lambda$ on a microscopic scale.

For the \emph{spatial} dependence of probabilities, one cannot expect an
analogous form, $p_i\sim\exp(\eta d i)$, if the system has spatial
disorder, i.~e. no translation symmetry. Instead, we expect that
\begin{equation}
p_i = p_0 C_i \exp(\eta d i),
\end{equation}
where the coefficients $C_i$ does not exponentially grow or decay when $i$
tends to infinity. Consequently,
\begin{equation}
\log \frac{p_i}{p_0} = \eta d i + O(1),
\end{equation}
which gives
\begin{equation}
\eta = \lim_{i\rightarrow\pm\infty} \frac1{di} \log \frac{p_i}{p_0},
\end{equation}
or, equivalently,
\begin{equation} \label{eq:2-mu-from-p}
\eta = \frac1d \left\langle \log \frac{p_{i+1}}{p_{i}} \right\rangle,
\end{equation}
where angle brackets denote averaging over the site number~$i$.

Eq.~(\ref{eq:2-mu-from-p}) can serve as the microscopic definition
of $\eta$. However, it is more convenient for our aim to define
$\eta$ in another way:
\begin{equation} \label{eq:2-mu-from-j}
\eta = \frac1d \left\langle \log \frac{j_{i,i+1}}{j_{i-1,i}} \right\rangle,
\end{equation}
where $j_{i,i+1}$ is the flow of carriers from site $i$ to site $i+1$:
\begin{equation} \label{eq:2-j-from-p}
j_{i,i+1} = \Gamma_{i,i+1}\,p_i - \Gamma_{i+1,i}\,p_{i+1}.
\end{equation}
It is easy to show that Eqs.~(\ref{eq:2-mu-from-p}) and
(\ref{eq:2-mu-from-j}) give equal values of $\eta$. Indeed,
in a macroscopic consideration the flow of particles $j(x,t)$ is
connected to the concentration $n(x,t)$ as
\begin{equation}
j = v\, n - D\, \partial n/\partial x.
\end{equation}
Therefore, if $n \sim\exp(\eta x)$ then $j \sim\exp(\eta x)$. Going to a
microscopic picture, one can get Eq.~(\ref{eq:2-mu-from-p}) from
$n\sim\exp(\eta x)$ and Eq.~(\ref{eq:2-mu-from-j}) from $j \sim\exp(\eta
x)$. Consequently the value of $\eta$ should be the same in all these
equations.

Let us now obtain $v$ and $D$ from Eq.~(\ref{eq:2-mu-from-j}). For
this purpose we rewrite Eq.~(\ref{eq:2-time-evolution-lambda})
taking into account Eq.~(\ref{eq:2-j-from-p}):
\begin{equation} \label{eq:2-time-evolution-lambda-j}
-\lambda p_i = j_{i-1,i} - j_{i,i+1},
\end{equation}
which gives
\begin{equation} \label{eq:2-ratio-j}
\frac{j_{i,i+1}}{j_{i-1,i}} = 1 + \lambda\frac{p_i}{j_{i-1,i}}.
\end{equation}
The ratio $p_i/j_{i-1,i}$ is a function of $\lambda$ since the
probability $p_i$ and the carrier flow $j_{i-1,i}$ are defined by a
$\lambda$-dependent equation~(\ref{eq:2-time-evolution-lambda}).
We expand this ratio in a Taylor series:
\begin{equation} \label{eq:2-ab-definition}
\frac{p_i}{j_{i-1,i}} = a_i + \lambda b_i + O(\lambda^2).
\end{equation}
(Our coefficients $a_i$ are the same as Derrida's $r_n$ in
Ref.~\onlinecite{Derrida1983}.) Substitution of
Eqs.~(\ref{eq:2-ratio-j}) and (\ref{eq:2-ab-definition}) into
Eq.~(\ref{eq:2-mu-from-j}) gives:

\begin{align}
\eta = {} & \frac1d \left\langle \log \left( 1 + \lambda\frac{p_i}{j_{i-1,i}} \right) \right\rangle = \notag \\
& \frac1d \left\langle \log \left( 1 + \lambda a_i + \lambda^2 b_i + O(\lambda^3) \right) \right\rangle = \\
& \frac{\lambda}{d} \langle a_i \rangle + \frac{\lambda^2}{d} \left( \langle b_i \rangle - \langle a_i^2 \rangle /2 \right) + O(\lambda^3). \notag
\end{align}

Comparing the latter equation with
Eq.~(\ref{eq:2-mu-from-lambda}), one obtains $v$ and $D$:
\begin{equation} \label{eq:2-vD-from-ab}
v = \frac{d}{\langle a_i \rangle}, \quad
D = d^2 \frac{\langle b_i \rangle - 
\langle a_i^2 \rangle /2 }{\langle a_i \rangle^3}.
\end{equation}
The expression for $v$ coincides with that obtained by Derrida
(Eq.~(63) of Ref.~\onlinecite{Derrida1983}), whereas the
expression for $D$ is a new result.

In the rest of this section, we obtain explicit expressions
for the quantities $a_i$ and $b_i$. The mean values $\langle a_i
\rangle$, $\langle a_i^2 \rangle$, and $\langle b_i \rangle$ will
be evaluated in Sec.~\ref{sec-analytic-RBM} for the RBM and in
Sec.~\ref{sec-analytic-REM} for the REM leading to the
analytical expressions for the diffusion coefficient $D$ in the
RBM and in the REM.

In order to find the coefficients $a_i$, we set $\lambda$ to zero in
Eq.~(\ref{eq:2-ab-definition}). As it is seen from
Eq.~(\ref{eq:2-time-evolution-lambda-j}), the carrier flow $j_{i,i+1}$
does not depend on $i$ in the case of $\lambda=0$. Dividing
Eq.~(\ref{eq:2-j-from-p}) by the carrier flow, one obtains a set of
equations for coefficients $a_i$:
\begin{equation} \label{eq:2-equation-for-a}
\forall i \quad \Gamma_{i,i+1}\,a_i - \Gamma_{i+1,i}\,a_{i+1} = 1.
\end{equation}
The solution of Eq.~(\ref{eq:2-equation-for-a}) can be presented as an
infinite series:
\begin{equation} \label{eq:2-solution-for-a}
a_i = \frac{1}{\Gamma_{i,i+1}} + 
\frac{\Gamma_{i+1,i}}{\Gamma_{i,i+1}\Gamma_{i+1,i+2}} +
\frac{\Gamma_{i+1,i}\Gamma_{i+2,i+1}}
{\Gamma_{i,i+1}\Gamma_{i+1,i+2}\Gamma_{i+2,i+3}} + \cdots,
\end{equation}
what can be checked directly by substituting
Eq.~(\ref{eq:2-solution-for-a}) into
Eq.~(\ref{eq:2-equation-for-a}). To prove the convergence of the
series~(\ref{eq:2-solution-for-a}), let us rewrite it using the
condition of detailed balance, Eq.~(\ref{eq:2-detailed-balance}):
\begin{equation} \label{eq:2-solution-for-a-energy}
\begin{array}{l}
a_i = \Gamma_{i,i+1}^{-1}
+ B^{-1}\exp\left(\frac{\varepsilon_{i+1}-\varepsilon_i}{kT}\right)\Gamma_{i+1,i+2}^{-1} + \\[2mm]
B^{-2}\exp\left(\frac{\varepsilon_{i+2}-\varepsilon_i}{kT}\right)\Gamma_{i+2,i+3}^{-1} + \cdots,
\end{array}
\end{equation}
where $B=\exp(eFd/kT)$. For any physically reasonable system, the
quantities
$\exp\left((\varepsilon_{i+k}-\varepsilon_i)/kT\right)\Gamma_{i+k,i+k+1}^{-1}$
can be regarded as having an upper boundary. Denoting this
boundary as $C$, we get an upper estimate for $a_i$:
\begin{equation}
a_i < C + B^{-1}C + B^{-2}C + \cdots = \frac{C}{1-B^{-1}}
\end{equation}
that proves convergence of the series~(\ref{eq:2-solution-for-a})
under the condition $B>1$, i.~e., $eF>0$.

In order to obtain $b_i$, we need a set of equations connecting
$b_i$ to $b_{i+1}$ in analogy with Eq.~(\ref{eq:2-equation-for-a})
that connects $a_i$ to $a_{i+1}$. We will derive the necessary
equations using Eq.~(\ref{eq:2-j-from-p}),
Eq.~(\ref{eq:2-ratio-j}), and the Taylor
expansion~(\ref{eq:2-ab-definition}). Let us first divide
Eq.~(\ref{eq:2-j-from-p}) by $j_{i,i+1}$ and slightly rearrange
it:
\begin{equation}
\Gamma_{i,i+1} \frac{j_{i-1,i}}{j_{i,i+1}}\, \frac{p_i}{j_{i-1,i}} - \Gamma_{i+1,i} \frac{p_{i+1}}{j_{i,i+1}} = 1.
\end{equation}
Let us now use the expansion~(\ref{eq:2-ab-definition}) for
quantities $p_i/j_{i-1,i}$:
\begin{equation} \label{eq:2-deduce-b-step1}
\Gamma_{i,i+1} \frac{j_{i-1,i}}{j_{i,i+1}} (a_i+\lambda b_i)
- \Gamma_{i+1,i} (a_{i+1}+\lambda b_{i+1}) = 1+O(\lambda^2).
\end{equation}
The latter equation contains the ratio $j_{i-1,i}/j_{i,i+1}$. We
derive this ratio from Eq.~(\ref{eq:2-ratio-j}) using also the
expansion~(\ref{eq:2-ab-definition}):
\begin{equation} \label{eq:2-deduce-b-step2}
\frac{j_{i-1,i}}{j_{i,i+1}}
= \frac{1}{1+\lambda p_i/j_{i-1,i}}
= \frac{1}{1+\lambda a_i+O(\lambda^2)}
= 1-\lambda a_i+O(\lambda^2).
\end{equation}
Finally, let us substitute Eq.~(\ref{eq:2-deduce-b-step2}) into
Eq.~(\ref{eq:2-deduce-b-step1}):
\begin{equation}
\Gamma_{i,i+1} (1-\lambda a_i) (a_i+\lambda b_i)
- \Gamma_{i+1,i} (a_{i+1}+\lambda b_{i+1}) = 1+O(\lambda^2),
\end{equation}
and collect separately terms, which do not contain $\lambda$,
and those proportional to $\lambda$. The former terms lead to
Eq.~(\ref{eq:2-equation-for-a}), while the latter ones give the
equation
\begin{equation} \label{eq:2-deduce-b-step4}
\Gamma_{i,i+1} (\lambda b_i-\lambda a_i^2) - \Gamma_{i+1,i} \lambda b_{i+1} = 0.
\end{equation}
Eq.~(\ref{eq:2-deduce-b-step4})
provides a desired set of equations for coefficients $b_i$:
\begin{equation} \label{eq:2-equation-for-b}
\forall i \quad \Gamma_{i,i+1}\,b_i - \Gamma_{i+1,i}\,b_{i+1} = \Gamma_{i,i+1}\,a_i^2.
\end{equation}

The solution of Eq.~(\ref{eq:2-equation-for-b}) can be found as an infinite series:
\begin{equation} \label{eq:2-solution-for-b}
b_i = a_i^2 + \frac{\Gamma_{i+1,i}}{\Gamma_{i,i+1}} a_{i+1}^2 +
  \frac{\Gamma_{i+1,i}\Gamma_{i+2,i+1}}{\Gamma_{i,i+1}\Gamma_{i+1,i+2}} a_{i+2}^2 + \cdots
\end{equation}
that can be checked by substitution into
Eq.~(\ref{eq:2-equation-for-b}). Like
Eq.~(\ref{eq:2-solution-for-a}), the
series~(\ref{eq:2-solution-for-b}) converges provided the product
$eF$ is positive. To prove it, we substitute the condition of
detailed balance, Eq.~(\ref{eq:2-detailed-balance}), into this
series:
\begin{equation} \label{eq:2-solution-for-b-energy}
\begin{array}{l}
b_i = a_i^2 + B^{-1}\exp\left(\frac{\varepsilon_{i+1}-\varepsilon_i}{kT}\right)a_{i+1}^2 + \\ B^{-2}\exp\left(\frac{\varepsilon_{i+2}-\varepsilon_i}{kT}\right)a_{i+2}^2 + \cdots
\end{array}
\end{equation}
In any real system we find an upper limit for the quantities
$\exp\left((\varepsilon_{i+k}-\varepsilon_i)/kT\right)a_{i+k}^2$.
Setting this limit equal to $\tilde C$, we
obtain an upper estimate for $b_i$:
\begin{equation}
b_i < \tilde C + B^{-1}\tilde C + B^{-2}\tilde C + \cdots 
= \frac{\tilde C}{1-B^{-1}}.
\end{equation}
Therefore the series~(\ref{eq:2-solution-for-b}) converges if $B>1$,
i.~e. if $eF>0$.

As a result, we have obtained an analytical
expression~(\ref{eq:2-vD-from-ab}) for the diffusion coefficient
$D$ in a one-dimensional hopping system. For coefficients $a_i$
and $b_i$ that contribute into Eq.~(\ref{eq:2-vD-from-ab}) we have
found series representations (\ref{eq:2-solution-for-a}) and
(\ref{eq:2-solution-for-b}) in the case $eF>0$. It is easy to
write down analogous series for $a_i$ and $b_i$ in the opposite
case, $eF<0$.

\section{Random-barrier model: exact results}
\label{sec-analytic-RBM} Let us now apply
Eqs.~(\ref{eq:2-vD-from-ab}), (\ref{eq:2-solution-for-a}), and
(\ref{eq:2-solution-for-b}) to the random-barrier model described
by Eqs.~(\ref{eq-gaussian-dos}) and (\ref{eq-rate-RBM}). In this
model, any two transition rates $\Gamma_{ij}$ and $\Gamma_{kl}$
are statistically independent, if $(ij)$ and $(kl)$ are different
pairs of sites. The rates $\Gamma_{i,i+1}$ and $\Gamma_{i+1,i}$,
related to the same pair are connected to each other. As a result,
all statistical properties of the random-barrier model are defined
by mean values $\langle \Gamma_{i,i+1}^m \Gamma_{i+1,i}^n \rangle$
with different $m$'s and $n$'s. We introduce the following
notations for these mean values:
\begin{equation} \label{eq:2-m-for-RBM}
\begin{array}{l}
m_1 = \langle \Gamma_{i+1,i} / \Gamma_{i,i+1} \rangle , \\[2mm]
m_2 = \langle \Gamma_{i+1,i}^2 / \Gamma_{i,i+1}^2 \rangle , \\[2mm]
m_3 = \langle 1 / \Gamma_{i,i+1} \rangle , \\[2mm]
m_4 = \langle 1 / \Gamma_{i,i+1}^2 \rangle , \\[2mm]
m_5 = \langle \Gamma_{i+1,i} / \Gamma_{i,i+1}^2 \rangle .
\end{array}
\end{equation}

In order to obtain the drift velocity $v$ and the diffusion
coefficient $D$ from Eq.~(\ref{eq:2-vD-from-ab}), one should
calculate the mean values $\langle a_i \rangle$, $\langle a_i^2
\rangle$, and $\langle b_i \rangle$. We start with calculating
$\langle a_i \rangle$. Let us denote successive terms in the
expansion~(\ref{eq:2-solution-for-a}) as $a^{(0)},\, a^{(1)},\,
a^{(2)},\, \ldots$ Then,
\begin{align}
\langle a^{(0)} \rangle &= 
\left\langle \frac{1}{\Gamma_{i,i+1}} \right\rangle = m_3, \notag \\
\langle a^{(1)} \rangle &= 
\left\langle \frac{\Gamma_{i+1,i}}{\Gamma_{i,i+1}} \right\rangle
 \, \left\langle \frac{1}{\Gamma_{i+1,i+2}} \right\rangle = m_1 m_3, \\
\langle a^{(2)} \rangle &= 
\left\langle \frac{\Gamma_{i+1,i}}{\Gamma_{i,i+1}} \right\rangle
 \, \left\langle \frac{\Gamma_{i+2,i+1}}{\Gamma_{i+1,i+2}} \right\rangle
 \, \left\langle \frac{1}{\Gamma_{i+2,i+3}} \right\rangle = m_1^2 m_3, \notag \\
\ldots & \notag \\
\langle a^{(k)} \rangle &= m_1^k m_3. \notag
\end{align}
Consequently,
\begin{equation} \label{eq:2-mean-a-RBM}
\begin{array}{l}
\langle a_i \rangle = \langle a^{(0)} \rangle + 
\langle a^{(1)} \rangle + \langle a^{(2)} \rangle + \ldots = \\[2mm]
m_3(1+m_1+m_1^2+\ldots) = m_3/(1-m_1).
\end{array}
\end{equation}

The mean value $\langle a_i^2 \rangle$ can be represented as a sum
of values $\langle a^{(k)} a^{(l)} \rangle$ over all pairs $k,l$:
\begin{equation} \label{eq:2-mean-a2-from-ak-al}
\langle a_i^2 \rangle = 
\sum_{k=0}^\infty \sum_{l=0}^\infty \langle a^{(k)} a^{(l)} \rangle.
\end{equation}
It is easy to check that
\begin{equation} \label{eq:2-ak-al-RBM}
\langle a^{(k)} a^{(l)} \rangle =
\left\{ \begin{array}{ll}
m_1^{l-k-1} m_2^k m_3 m_5, & \mbox{if } k<l, \\[2mm]
m_2^k m_4,                 & \mbox{if } k=l, \\[2mm]
m_1^{k-l-1} m_2^l m_3 m_5, & \mbox{if } k>l.
\end{array} \right.
\end{equation}
Then, presenting Eq.~(\ref{eq:2-mean-a2-from-ak-al}) in the form
\begin{equation}
\langle a_i^2 \rangle = \sum_{k=0}^\infty \langle (a^{(k)})^2 \rangle
+ 2 \sum_{k<l} \langle a^{(k)} a^{(l)} \rangle,
\end{equation}
using Eq.~(\ref{eq:2-ak-al-RBM}) and introducing the notation $p=l-k-1$, we obtain
\begin{equation} \label{eq:2-mean-a2-RBM}
\begin{array}{l}
\langle a_i^2 \rangle = \sum\limits_{k=0}^\infty m_2^k m_4 
+ 2 \sum\limits_{k=0}^\infty \sum\limits_{l=k+1}^\infty m_1^{l-k-1} m_2^k m_3 m_5 
= \\[4mm]
\frac{m_4}{1-m_2} 
+ 2m_3m_5 \sum\limits_{k=0}^\infty m_2^k \sum\limits_{p=0}^\infty m_1^p 
= \frac{m_4}{1-m_2} + \frac{2m_3m_5}{(1-m_1)(1-m_2)}.
\end{array}
\end{equation}

In an analogous way, we denote successive terms of the
series~(\ref{eq:2-solution-for-b}) as $b^{(0)},\, b^{(1)},\,
b^{(2)},\, \ldots$ The mean values of these quantities are
\begin{align}
\langle b^{(0)} \rangle &= \langle a_i^2 \rangle, \notag \\
\langle b^{(1)} \rangle 
&= \left\langle \frac{\Gamma_{i+1,i}}{\Gamma_{i,i+1}} a_{i+1}^2 \right\rangle  
= m_1 \langle a_i^2 \rangle, \notag \\
\langle b^{(2)} \rangle &= 
\left\langle \frac{\Gamma_{i+1,i}\Gamma_{i+2,i+1}}
{\Gamma_{i,i+1}\Gamma_{i+1,i+2}} a_{i+2}^2 \right\rangle  
= m_1^2 \langle a_i^2 \rangle,\\
\ldots & \notag \\
\langle b^{(k)} \rangle &= m_1^k \langle a_i^2 \rangle. \notag
\end{align}
Therefore,
\begin{equation} \label{eq:2-mean-b-RBM}
\begin{array}{l}
\langle b_i \rangle = \langle b^{(0)} \rangle + \langle b^{(1)} \rangle + \langle b^{(2)} \rangle + \ldots = \\[2mm]
\langle a_i^2 \rangle\, (1+m_1+m_1^2+\ldots) = \langle a_i^2 \rangle/(1-m_1).
\end{array}
\end{equation}

Finally, we substitute  Eqs.~(\ref{eq:2-mean-a-RBM}),
(\ref{eq:2-mean-a2-RBM}), and (\ref{eq:2-mean-b-RBM}) for the mean
values $\langle a_i \rangle$, $\langle a_i^2 \rangle$ and $\langle
b_i \rangle$ into Eq.~(\ref{eq:2-vD-from-ab}). This gives
\begin{equation} \label{eq:2-v-from-m-RBM}
v = d\, \frac{1-m_1}{m_3},
\end{equation}
\begin{equation} \label{eq:2-D-from-m-RBM}
D = d^2 \frac{1-m_1^2}{2m_3^3(1-m_2)}\, (m_4(1-m_1) + 2m_3m_5).
\end{equation}
Equations~(\ref{eq:2-v-from-m-RBM}) and~(\ref{eq:2-D-from-m-RBM})
are not new results---they were obtained in
Ref.~\onlinecite{Derrida1983} (Eq.~(67) and Eq.~(70),
respectively). Their derivation in the frame of our method clearly
demonstrates that both methods (Derrida's and ours) are
consistent. In the rest of this subsection, we will apply these
equations to the case of Gaussian distribution of barrier heights.

For the transition rates defined by Eqs.~(\ref{eq-gaussian-dos}) and
(\ref{eq-rate-RBM}), the mean values $m_1 \ldots m_5$ are easy to
evaluate. Setting $\Gamma_0$ equal to unity for the sake of
simplicity, we obtain
\begin{equation} \label{eq:2-m-answer-RBM}
\begin{array}{l}
m_1 = \exp\left( - \frac{eFd}{kT} \right), \\[2mm]
m_2 = \exp\left( - \frac{2eFd}{kT} \right), \\[2mm]
m_3 = \exp\left( \frac{\sigma^2}{8(kT)^2} - \frac{eFd}{2kT} \right), \\[2mm]
m_4 = \exp\left( \frac{\sigma^2}{2(kT)^2} - \frac{eFd}{kT} \right), \\[2mm]
m_5 = \exp\left( \frac{\sigma^2}{8(kT)^2} - \frac{3eFd}{2kT} \right).
\end{array}
\end{equation}
From Eq.~(\ref{eq:2-v-from-m-RBM}) one obtains the result for the
drift velocity $v$:
\begin{equation} \label{eq:2-v-RBM}
v = 2d \exp\left( -\frac{\sigma^2}{8(kT)^2} \right) \sinh\left( \frac{eFd}{2kT} \right).
\end{equation}
Note that we derived the latter equation only for the case $eF>0$.
However, it is easy to show that Eq.~(\ref{eq:2-v-RBM}) is valid
for any direction of the electric field. Indeed, the right-hand side of the
equation is an odd function of the electric field $F$. The
left-hand side (drift velocity) should also be odd, because the
system is symmetrical with respect to a left-to-right mirror
reflection $(x\rightarrow-x,\, F\rightarrow-F,\, v\rightarrow-v)$.
Therefore, if Eq.~(\ref{eq:2-v-RBM}) is satisfied for positive
electric fields, it remains valid for negative fields, and vice
versa.

An expression for the diffusion coefficient $D$ as a function of
$F$ can be obtained by substituting the mean
values~(\ref{eq:2-m-answer-RBM}) into
Eq.~(\ref{eq:2-D-from-m-RBM}). Strictly speaking this procedure is
valid for $D$ only in the case $eF>0$. One can however generalize
this expression for any sign of the electric field using the fact
that (for symmetry reasons) $D$ is an even function of $F$. One
simply should replace $eF$ by its absolute value, $|eF|$ in all
expressions. The result reads:
\begin{eqnarray} \label{eq:2-D-RBM}
\nonumber D = d^2 \exp\left( -\frac{\sigma^2}{8(kT)^2} 
- \frac{|eF|d}{2kT} \right) + \\
d^2 \exp\left( \frac{\sigma^2}{8(kT)^2} \right) 
\sinh\left( \frac{|eF|d}{2kT} \right).
\end{eqnarray}
Eq.~(\ref{eq:2-D-RBM}) was obtained for \emph{non-zero} electric
fields. However, one can check that it holds also for $F=0$.

Eq.~(\ref{eq:2-D-RBM}) differs from the expression given by
Bouchaud and Georges,\cite{Bouchaud1989}, $D(F) - D(0) \propto F
\exp[3 \sigma^2 / 8(kT)^2]$, though it is linear in $F$ to first
order. Eq.~(\ref{eq:2-D-RBM}) is plotted in Fig.~\ref{fig-1d-RB},
together with numerical results obtained in
Sec.~\ref{sec-numeric}.

\section{Random-energy model: exact results}
\label{sec-analytic-REM} The random-energy model in one dimension
implies the following definition of transition rates:
\begin{equation} \label{eq:2-MA-rate-REM}
 \Gamma_{i,i\pm1} = \Gamma_0 \exp\left( -\frac{\Delta\varepsilon_{i,i\pm1}+|\Delta\varepsilon_{i,i\pm1}|}{2kT} \right),
\end{equation}
where $\Delta\varepsilon_{i,i\pm1} =
\varepsilon_{i\pm1}-\varepsilon_i\mp eFd$ is the difference
between the energies of a charge carrier on the final site and on
the initial site, respectively for each jump. For simplicity, we
set the constant $\Gamma_0\equiv\nu_0 \exp(-2d/a)$ to unity.

For the REM, one can use the same way of calculating the velocity
and the diffusion constant as for the RBM. 
The REM contains more correlations between transition rates that the RBM, 
which leads to more complicated calculations of the mean
values $\langle a_i \rangle$, $\langle a_i^2 \rangle$, and
$\langle b_i \rangle$. In the REM, each rate $\Gamma_{ij}$ depends
on the energies $\varepsilon_i$ and $\varepsilon_j$, which are
independent random variables. Therefore, the rates $\Gamma_{ij}$
and $\Gamma_{kl}$ are correlated if the pairs of sites $(ij)$ and
$(kl)$ have at least one site in common.

We will see below that the drift velocity $v$ and the diffusion
coefficient $D$ depend on eleven quantities $m_1\ldots m_{11}$
related to the statistics of site energies and transition rates:

\begin{align}
m_1 = {} & \langle e^{-\varepsilon_i/kT} \rangle , \notag \\
m_2 = {} & \langle e^{-2\varepsilon_i/kT} \rangle , \notag \\
m_3 = {} & \langle \Gamma_{i,i+1}^{-1} \rangle , \notag \\
m_4 = {} & \langle e^{\varepsilon_i/kT}\, \Gamma_{i,i+1}^{-1} \rangle , \notag \\
m_5 = {} & \langle e^{-\varepsilon_i/kT}\, \Gamma_{i,i+1}^{-1} \rangle , \notag \\
m_6 = {} & \langle \Gamma_{i,i+1}^{-2} \rangle , \\
m_7 = {} & \langle e^{\varepsilon_i/kT}\, \Gamma_{i,i+1}^{-2} \rangle , \notag \\
m_8 = {} & \langle e^{2\varepsilon_i/kT}\, \Gamma_{i,i+1}^{-2} \rangle , \notag \\
m_9 = {} & \langle e^{(\varepsilon_{i+1}-\varepsilon_i)/kT}\, \Gamma_{i,i+1}^{-1}\, \Gamma_{i+1,i+2}^{-1} \rangle , \notag \\
m_{10} = {} & \langle e^{\varepsilon_{i+1}/kT}\, \Gamma_{i,i+1}^{-1}\, \Gamma_{i+1,i+2}^{-1} \rangle , \notag \\
m_{11} = {} & \langle e^{(\varepsilon_{i+1}+\varepsilon_i)/kT}\, \Gamma_{i,i+1}^{-1}\, \Gamma_{i+1,i+2}^{-1} \rangle . \notag
\end{align}

In the following we proceed for the REM along the same steps 
as for the RBM in the previous section.

\subsubsection{Calculation of $\langle a_i \rangle$}

Let us denote successive terms of the
expansion~(\ref{eq:2-solution-for-a-energy}) as $a^{(0)}$,
$a^{(1)}$, $a^{(2)}$, and so on. Then $\langle a_i \rangle =
\langle a^{(0)} \rangle + \langle a^{(1)} \rangle + \langle
a^{(2)} \rangle + \ldots$ . The latter quantities can be easily
expressed via $m_1$, $m_3$ and $m_4$:
 \begin{align}
  \langle a^{(0)} \rangle &= \langle \Gamma_{i,i+1}^{-1} \rangle = m_3,   \\
  \langle a^{(1)} \rangle &= B^{-1} \langle e^{-\varepsilon_i/kT} \rangle \langle e^{\varepsilon_{i+1}/kT}\Gamma_{i+1,i+2}^{-1} \rangle = B^{-1} m_1 m_4,    \notag \\
  \langle a^{(2)} \rangle &= B^{-2} \langle e^{-\varepsilon_i/kT} \rangle \langle e^{\varepsilon_{i+2}/kT}\Gamma_{i+2,i+3}^{-1} \rangle = B^{-2} m_1 m_4, \notag
 \end{align}

and, generally, $\langle a^{(k)} \rangle = B^{-k} m_1 m_4$ for any $k>0$. Then,
\begin{equation} \label{eq:2-REM-ai}
\langle a_i \rangle = m_3 + (B^{-1}+B^{-2}+\ldots) m_1 m_4 = m_3+\frac{m_1m_4}{B-1}.
\end{equation}

\subsubsection{Calculation of $\langle a_i^2 \rangle$}

According to Eq.~(\ref{eq:2-mean-a2-from-ak-al}), the calculation
of $\langle a_i^2 \rangle$ is reduced calculating the mean values
$\langle a^{(k)} a^{(l)} \rangle$ for all integer $k\geq 0$ and
$l\geq 0$. Thus one can reduce the mean values $\langle a^{(k)} a^{(l)}
\rangle$ to:
\begin{equation} \label{eq:2-ak-al-REM}
\langle a^{(k)} a^{(l)} \rangle =
\left\{ \begin{array}{ll}
m_6,                 & \mbox{if } k=l=0, \\[2mm]
B^{-1} m_9,          & \mbox{if } k=0,\, l=1, \\[2mm]
B^{-l} m_4 m_5,      & \mbox{if } k=0,\, l>1, \\[2mm]
B^{-2k} m_2 m_8,     & \mbox{if } k>0,\, l=k, \\[2mm]
B^{-2k-1} m_2 m_{11},& \mbox{if } k>0,\, l=k+1, \\[2mm]
B^{-k-l} m_2 m_4^2,  & \mbox{if } k>0,\, l>k+1.
\end{array} \right.
\end{equation}
The next step is the estimate of the infinite
series~(\ref{eq:2-mean-a2-from-ak-al}). It is
convenient to rearrange the summation in
Eq.~(\ref{eq:2-mean-a2-from-ak-al}), separating terms
corresponding to different lines of Eq.~(\ref{eq:2-ak-al-REM}):
\begin{equation} \label{eq:2-mean-a2-from-ak-al-REM}
\begin{array}{l}
\langle a_i^2 \rangle = \langle (a^{(0)})^2 \rangle + \sum\limits_{k=1}^\infty \langle (a^{(k)})^2 \rangle + \\[3mm]
2 \langle a^{(0)} a^{(1)} \rangle + 2 \sum\limits_{l=2}^\infty \langle a^{(0)} a^{(l)} \rangle + \\[3mm]
2 \sum\limits_{k=1}^\infty \langle a^{(k)} a^{(k+1)} \rangle +
2 \sum\limits_{k=1}^\infty \sum\limits_{p=2}^\infty  \langle a^{(k)} a^{(k+p)} \rangle,
\end{array}
\end{equation}
where $p=l-k$. Substituting Eq.~(\ref{eq:2-ak-al-REM}) into this expansion, one gets:
\begin{align}
\langle a_i^2 \rangle = {} &  m_6 + m_2m_8 \sum\limits_{k=1}^\infty B^{-2k} + 2B^{-1}m_9 + \notag \\
&2m_4m_5 \sum\limits_{l=2}^\infty B^{-l} + 2m_2m_{11} \sum\limits_{k=1}^\infty B^{-2k-1} + \\
&2m_2m_4^2 \sum\limits_{k=1}^\infty B^{-2k} \sum\limits_{p=2}^\infty  B^{-p}. \notag
\end{align}

Finally, one can sum up the geometric series. The result reads:
\begin{equation} \label{eq:2-REM-ai2}
\begin{array}{l}
\langle a_i^2 \rangle = m_6 + \frac{2m_9}{B} + \frac{2m_4m_5}{B(B-1)} + \\[3mm]
m_2\frac{Bm_8+2m_{11}}{B(B^2-1)} + \frac{2m_2m_4^2}{B(B-1)(B^2-1)}.
\end{array}
\end{equation}

\subsubsection{Calculation of $\langle b_i \rangle$}

In an analogous way, $\langle b_i \rangle$ can be expressed as a
sum $\langle b^{(0)} \rangle + \langle b^{(1)} \rangle + \langle
b^{(2)} \rangle + \ldots$, where $b^{(0)}$, $b^{(1)}$...
$b^{(i)}$, ... are the terms of the
expansion~(\ref{eq:2-solution-for-b-energy}). Keeping in mind
that, according to Eq.~(\ref{eq:2-solution-for-a}), the values
$a_{i+1},\,a_{i+2},\,\ldots$ do not depend on $\varepsilon_i$, one
can express the mean values $b^{(k)}$ as follows:
\begin{align}
  \langle b^{(0)} \rangle &= \langle a_i^2 \rangle,   \notag \\
  \langle b^{(1)} \rangle &= B^{-1} \langle e^{-\varepsilon_i/kT} \rangle \langle e^{\varepsilon_{i+1}/kT}a_{i+1}^2 \rangle = B^{-1} m_1 M,    \\
  \langle b^{(2)} \rangle &= B^{-2} \langle e^{-\varepsilon_i/kT} \rangle \langle e^{\varepsilon_{i+2}/kT}a_{i+2}^2 \rangle = B^{-2} m_1 M,   \notag  \\
 \end{align}

and so on for larger $i$, where $M=\langle
e^{\varepsilon_i/kT}a_i^2 \rangle$. Thus,
\begin{equation}
  \langle b_i \rangle = \langle a_i^2 \rangle + (B^{-1}+B^{-2}+\ldots) m_1 M,
\end{equation}
or
\begin{equation} \label{eq:2-bi-from-M-REM}
  \langle b_i \rangle = \langle a_i^2 \rangle + \frac{m_1}{B-1} M.
\end{equation}

In order to find $M$, one can expand it in series analogous to Eq.~(\ref{eq:2-mean-a2-from-ak-al}):
\begin{equation}
M \equiv \langle e^{\varepsilon_i/kT} a_i^2 \rangle = \sum_{k=0}^\infty \sum_{l=0}^\infty \langle e^{\varepsilon_i/kT} a^{(k)} a^{(l)} \rangle,
\end{equation}
and express each term of the expansion via:
\begin{equation} \label{eq:2-eps-ak-al-REM}
\langle e^{\frac{\varepsilon_i}{kT}} a^{(k)} a^{(l)} \rangle =
\left\{ \begin{array}{ll}
m_7,                 & \mbox{if } k=l=0, \\[2mm]
B^{-1} m_{10},       & \mbox{if } k=0,\, l=1, \\[2mm]
B^{-l} m_4 m_3,      & \mbox{if } k=0,\, l>1, \\[2mm]
B^{-2k} m_1 m_8,     & \mbox{if } k>0,\, l=k, \\[2mm]
B^{-2k-1} m_1 m_{11},& \mbox{if } k>0,\, l=k+1, \\[2mm]
B^{-k-l} m_1 m_4^2,  & \mbox{if } k>0,\, l>k+1.
\end{array} \right.
\end{equation}

The following steps are the same as the ones leading from
Eq.~(\ref{eq:2-ak-al-REM}) to Eq.~(\ref{eq:2-REM-ai2}). Instead of
proceeding in this way, one can recognize that
Eq.~(\ref{eq:2-ak-al-REM}) transforms to
Eq.~(\ref{eq:2-eps-ak-al-REM}) by the following replacements:
\[
  m_6 \rightarrow m_7, \quad m_9 \rightarrow m_{10}, \quad m_5 \rightarrow m_3, \quad m_2 \rightarrow m_1.
\]
Applying the same replacements to
Eq.~(\ref{eq:2-REM-ai2}), we obtain the result for $M$:
\begin{equation} \label{eq:2-REM-M}
\begin{array}{l}
M = m_7 + \frac{2m_{10}}{B} + \frac{2m_3m_4}{B(B-1)} + \\[3mm]
m_1\frac{Bm_8+2m_{11}}{B(B^2-1)} + \frac{2m_1m_4^2}{B(B-1)(B^2-1)}.
\end{array}
\end{equation}
Substituting Eqs.~(\ref{eq:2-REM-ai2}) and (\ref{eq:2-REM-M}) into
Eq.~(\ref{eq:2-bi-from-M-REM}), one obtains the expression for
$\langle b_i \rangle$ in terms of the values $m_1\ldots m_{11}$.

\subsubsection{Drift velocity and diffusion coefficient}

Combining equations~(\ref{eq:2-vD-from-ab}), (\ref{eq:2-REM-ai}),
(\ref{eq:2-REM-ai2}), (\ref{eq:2-bi-from-M-REM}),
(\ref{eq:2-REM-M}) leads to
\begin{widetext}
\begin{equation} \label{eq:2-v-from-m-REM}
v = \frac{d}{m_3+m_1m_4/(B-1)},
\end{equation}
\begin{eqnarray} \label{eq:2-D-from-m-REM}
\nonumber D =   %  \frac{d^2}{2 \langle a_i \rangle^3} \left( \langle a_i^2 \rangle + \frac{2m_1}{B-1}M \right) =
\frac{v^3}{2d} \left[ m_6 + \frac{2m_9}{B} + \frac{2m_4m_5}{B(B-1)} + m_2\frac{Bm_8+2m_{11}}{B(B^2-1)} + \frac{2m_2m_4^2}{B(B-1)(B^2-1)} + \right. \\
\left. \frac{2m_1}{B-1} \left( m_7 + \frac{2m_{10}}{B} + \frac{2m_3m_4}{B(B-1)} + m_1\frac{Bm_8+2m_{11}}{B(B^2-1)} + \frac{2m_1m_4^2}{B(B-1)(B^2-1)} \right) \right],
\end{eqnarray}
\end{widetext}
where $B=\exp(eFd/kT)$. Note that we derived these equations
for the case $eF>0$. One can easily generalize the equations for
the case $eF<0$, keeping in mind that $v$~is an odd function of
$F$, and $D$~is an even function.

Eq.~(\ref{eq:2-v-from-m-REM}) was obtained previously by Cordes
\emph{et al.},\cite{Cordes2001} using Derrida's
method,\cite{Derrida1983} while Eq.~\ref{eq:2-D-from-m-REM} is a new result.

Equations~(\ref{eq:2-v-from-m-REM}), (\ref{eq:2-D-from-m-REM}) are
general for the case of the REM---their derivation is not
restricted by a special choice of the density of states or by the
choice of the relation between the transition rates $\Gamma_{ij}$
and the site energies $\varepsilon_i,\,\varepsilon_j$. We used
only four assumptions: (i) all sites are arranged in the line with
constant distance $d$ between them; (ii) there are only
transitions between nearest neighbors; (iii) transition rates
``forth'' and ``back'' ($\Gamma_{i,i+1}$ and $\Gamma_{i+1,i}$)
obey the principle of detailed balance,
Eq.~(\ref{eq:2-detailed-balance}); (iv) energies of different
sites are independent random variables having the same
distribution function.

\subsubsection{Gaussian density of states}

\begin{table*} 
 \caption{The values of $m_1\ldots m_{11}$ for the random-energy model
with Gaussian density of states. ``MA'' refers to using Miller-Abrahams
hopping rates, Eq.~(\ref{eq:2-MA-rate-REM}), ``modified MA''---to
hopping rates defined by Eq.~(\ref{eq:2-modified-MA-rate-REM}).
Other notations: $A=\exp(\sigma^2/(kT)^2)$, $B=\exp(eFd/kT)$, 
$\alpha=\sigma/kT$,
$\beta=eFd/\sigma$; erfc is the complementary error function; the function
$\mathcal{F}(a,b)$ is defined by Eq.~(\ref{eq:2-Fab-def}).
}\label{tab:2-m}
 \begin{tabular}{|c|c|c|c|} \hline
Notation & Definition & Value (MA) & Value (modified MA) \\ \hline\hline
$m_1$ & $\langle e^{-\varepsilon_i/kT} \rangle$                       & $\sqrt A$ & $\sqrt A$ \\ \hline
$m_2$ & $\langle e^{-2\varepsilon_i/kT} \rangle$                      & $A^2$     & $A^2$     \\ \hline
$m_3$ & $\langle \Gamma_{i,i+1}^{-1} \rangle$                         &
   $\frac12[$erfc$(-\frac\beta2)+AB^{-1}$erfc$(\frac\beta2-\alpha)]$  & $1+AB^{-1}$  \\ \hline
$m_4$ & $\langle e^{\varepsilon_i/kT}\, \Gamma_{i,i+1}^{-1} \rangle$  &
   $\frac12\sqrt A[$erfc$(-\frac\beta2-\frac\alpha2)+B^{-1}$erfc$(\frac\beta2-\frac\alpha2)]$  & $\sqrt A(1+B^{-1})$  \\ \hline
$m_5$ & $\langle e^{-\varepsilon_i/kT}\, \Gamma_{i,i+1}^{-1} \rangle$ &
   $\frac12\sqrt A[$erfc$(-\frac\beta2+\frac\alpha2)+A^2B^{-1}$erfc$(\frac\beta2-\frac{3\alpha}2)]$  & $\sqrt A(1+A^2B^{-1})$  \\ \hline
$m_6$ & $\langle \Gamma_{i,i+1}^{-2} \rangle$                         &
   $\frac12[$erfc$(-\frac\beta2)+A^4B^{-2}$erfc$(\frac\beta2-2\alpha)]$  & $1+2AB^{-1}+A^4B^{-2}$  \\ \hline
$m_7$ & $\langle e^{\varepsilon_i/kT}\, \Gamma_{i,i+1}^{-2} \rangle$  &
   $\frac12\sqrt A[$erfc$(-\frac\beta2-\frac\alpha2)+A^2B^{-2}$erfc$(\frac\beta2-\frac{3\alpha}2)]$  & $\sqrt A(1+2B^{-1}+A^2B^{-2})$  \\ \hline
$m_8$ & $\langle e^{2\varepsilon_i/kT}\, \Gamma_{i,i+1}^{-2} \rangle$ &
   $\frac12 A^2[$erfc$(-\frac\beta2-\alpha)+B^{-2}$erfc$(\frac\beta2-\alpha)]$  & $A^2(1+B^{-2})+2AB^{-1}$  \\ \hline
&& $A\mathcal{F}(-3\beta,2\alpha-\beta)+A^3B^{-2}\mathcal{F}(3\beta-3\alpha,\beta-3\alpha)$ & \\
   $m_9$ & $\langle e^{(\varepsilon_{i+1}-\varepsilon_i)/kT}\, \Gamma_{i,i+1}^{-1}\, \Gamma_{i+1,i+2}^{-1} \rangle$ &
   $+AB^{-1}[\frac12$erfc$(\frac{\alpha-\beta}{2})-\mathcal{F}(3\alpha-3\beta,\alpha-\beta)]$
   & $A(1+A^2B^{-2}+(1+A^3)B^{-1})$ \\
   && $+A^4B^{-1}[\frac12$erfc$(\frac\beta2-2\alpha)-\mathcal{F}(3\beta,\beta-4\alpha)]$ &  \\ \hline
&& $\sqrt A\mathcal{F}(-\alpha\!-\!3\beta,\alpha\!-\!\beta)+A^{3/2}B^{-2}\mathcal{F}(3\beta\!-\!2\alpha,\beta\!-\!2\alpha)$ & \\
   $m_{10}$ & $\langle e^{\varepsilon_{i+1}/kT}\, \Gamma_{i,i+1}^{-1}\, \Gamma_{i+1,i+2}^{-1} \rangle$ &
   $+\sqrt AB^{-1}[\frac12$erfc$(-\frac{\beta}{2})-\mathcal{F}(2\alpha-3\beta,-\beta)]$ &
   $\sqrt A(1+AB^{-2}+(1+A^2)B^{-1})$ \\
   && $+A^{5/2}B^{-1}[\frac12$erfc$(\frac\beta2-\frac{3\alpha}2)-\mathcal{F}(\alpha+3\beta,\beta-3\alpha)]$ & \\ \hline
&& $A\mathcal{F}(-2\alpha-3\beta,-\beta)+AB^{-2}\mathcal{F}(3\beta-\alpha,\beta-\alpha)$ & \\
   $m_{11}$ & $\langle e^{(\varepsilon_{i+1}+\varepsilon_i)/kT}\,\Gamma_{i,i+1}^{-1}\,\Gamma_{i+1,i+2}^{-1}\rangle$ &
   $+AB^{-1}[\frac12$erfc$(\frac{-\alpha-\beta}{2})-\mathcal{F}(\alpha-3\beta,-\alpha-\beta)]$ &
   $A(1+B^{-2}+(1+A)B^{-1})$ \\
   && $+A^2B^{-1}[\frac12$erfc$(\frac\beta2-\alpha)-\mathcal{F}(2\alpha+3\beta,\beta-2\alpha)]$ & \\ \hline
\end{tabular}
\end{table*}

We will now evaluate the quantities $m_1\ldots m_{11}$, assuming a
Gaussian density of states, Eq.~(\ref{eq-gaussian-dos}), and the
Miller-Abrahams transition rates, Eq.~(\ref{eq:2-MA-rate-REM}).
This evaluation is straightforward, for example:
\[
\begin{array}{l}
 m_1 = N_0^{-1} \int e^{-\varepsilon/kT} g(\varepsilon) d\varepsilon = \\[3mm]
\dfrac{1}{\sqrt{2\pi}\sigma} \int \exp\left(-\frac{\varepsilon}{kT}-\frac{\varepsilon^2}{2\sigma^2}\right) d\varepsilon =
\exp\left(\frac{\sigma^2}{2(kT)^2}\right); \\[3mm]
m_3 = N_0^{-2} \iint \Gamma_{12}^{-1} g(\varepsilon_1) g(\varepsilon_2) d\varepsilon_1 d\varepsilon_2 = \\[3mm]
\dfrac{1}{2\pi\sigma^2} \iint \exp\left(\frac{\varepsilon_2-\varepsilon_1-eFd+|\varepsilon_2-\varepsilon_1-eFd|}{2kT}-\frac{\varepsilon_1^2+\varepsilon_2^2}{2\sigma^2}\right) d\varepsilon_1 d\varepsilon_2.
\end{array}
\]
The latter integral can be evaluated by substitution
$u=\varepsilon_1+\varepsilon_2$, $v=\varepsilon_1-\varepsilon_2$, and
the result is
\begin{align}
 m_3 = {} &  \frac12 \mathrm{erfc}\left(-\frac{eFd}{2\sigma}\right) + \notag \\
&\frac12 \exp\left(\frac{\sigma^2}{(kT)^2}-\frac{eFd}{kT}\right) \mathrm{erfc}\left(\frac{eFd}{2\sigma}-\frac{\sigma}{kT}\right),
\end{align}
where erfc is the complementary error function, erfc$(x) = 2\pi^{-1/2}
\int_x^\infty e^{-t^2}dt = 1-$erf$(x)$.

\begin{figure}
\includegraphics[scale=0.5]{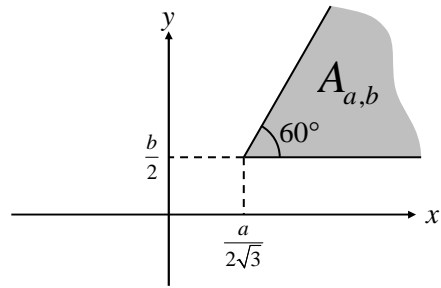}
\caption{\label{fig:F_def} Integration area for the definition of the function $\mathcal{F}(a,b)$.}
\end{figure}

The values $m_9\ldots m_{11}$ are triple integrals. They cannot be
expressed in elementary functions, but they can be reduced (by a
substitution $u=\varepsilon_1+\varepsilon_2+\varepsilon_3$,
$v=\varepsilon_1-\varepsilon_2$,
$w=\varepsilon_1+\varepsilon_2-2\varepsilon_3$) to a function
$\mathcal{F}(a,b)$ defined as
\begin{equation} \label{eq:2-Fab-def}
  \mathcal{F}(a,b) = \frac{1}{\pi} \iint\limits_{A_{a,b}} e^{-(x^2+y^2)}\, dx\, dy,
\end{equation}
where the area of integration $A_{a,b}$ is shown in Fig.~\ref{fig:F_def}.

The results are collected in Table~\ref{tab:2-m}. These results obtained for
Miller-Abrahams hopping rates, Eq.~(\ref{eq:2-MA-rate-REM}) look
rather complicated for analytical estimates. Below in Sec.~\ref{sec-numeric}
we use these expressions from Table~\ref{tab:2-m} for numerical calculations
and present the results for the Miller-Abrahams hopping rates,
Eq.~(\ref{eq:2-MA-rate-REM}). Here we will proceed with analytical
calculations based on slightly modified expressions for the
hopping rates, which allow straightforward analytical estimates.
We suggest to use, instead of Eq.~(\ref{eq:2-MA-rate-REM}), the
following ``modified Miller-Abrahams rates'' :
\begin{equation} \label{eq:2-modified-MA-rate-REM}
 \Gamma_{i,i\pm1} = \Gamma_0 \left[ 1 
+ \exp\left( \frac{\Delta\varepsilon_{i,i\pm1}}{kT} \right) \right]^{-1},
\end{equation}
where the constant $\Gamma_0$ will be set equal to unity for
the sake of simplicity. The difference between
Eq.~(\ref{eq:2-MA-rate-REM}) and
Eq.~(\ref{eq:2-modified-MA-rate-REM}) becomes negligible when
$|\Delta\varepsilon_{i,i\pm1}|\gg kT$. Therefore, for $\sigma\gg
kT$ we expect a good agreement between results obtained with these
two kinds of hopping rates.

Results for the ``modified Miller-Abrahams rates'' are also shown
in Table~\ref{tab:2-m}. Substituting them into Eqs.~(\ref{eq:2-v-from-m-REM}),
(\ref{eq:2-D-from-m-REM}), one can get the explicit expressions
for the drift velocity and the diffusion coefficient: \footnote{We
replace here $v$ and $eF$ by their absolute values, in
  order to not restrict ourselves to the case $eF>0$.}
\begin{widetext}
\begin{equation} \label{eq:2-v-modMA-REM}
|v| = d \, \frac{B-1}{2A+B-1},
\end{equation}
\begin{equation} \label{eq:2-D-modMA-REM}
D = d^2 \, \frac{ 4A^4(B-1)^2+16A^3(B-1)+16A^2+2A(3B+1)(B^2-1)+(B-1)^3(B+1)}{2(B+1)(2A+B-1)^3} ,
\end{equation}
\end{widetext}
where $A=\exp(\sigma^2/(kT)^2)$, $B=\exp(|eF|d/kT)$.

Now we will consider the mobility $\mu(F)=v(F)/F$ and the
diffusion coefficient $D(F)$ in the limit of small field. We
restrict ourselves to the case of modified Miller-Abrahams rates.
For $F=0$, one can obtain from Eqs.~(\ref{eq:2-v-modMA-REM}) and
(\ref{eq:2-D-modMA-REM}):
\begin{equation}
  D(0) = \mu(0)\,\frac{kT}{e} = \frac{d^2}2\exp(-(\sigma/kT)^2) .
\end{equation}

For small temperatures, $kT\ll\sigma$, the mobility and the
diffusion coefficient can be approximated by simple expressions:
\begin{equation} \label{eq:2-approx_mu}
  \mu(F) \approx \frac{ed^2}{2AkT}+\frac{|Fe|ed^3}{4A(kT)^2}+\frac{F^2e^3d^4}{12A(kT)^3} \,,
\end{equation}
\begin{equation} \label{eq:2-approx_D}
  D(F) \approx \frac{d^2}{2A}+\frac{|Fe|d^3}{2kT}+\frac{F^2e^2d^4A}{8(kT)^2}+\frac{|Fe|^3d^5A}{16(kT)^3} \,.
\end{equation}
These approximations are valid for sufficiently small fields, $|eFd|<kT$.

From the latter approximated expressions it is obvious that there is a
cusp at $F=0$ for both $\mu(F)$ and $D(F)$. The dependence $D(F)$
demonstrates a linear behavior for very small fields $(|eFd| \ll
kT/A)$, when the first two terms in Eq.~(\ref{eq:2-approx_D}) are
dominating, and a parabolic behavior for intermediate fields $(kT/A \ll
|eFd| \ll kT)$, when the third term is dominating.

\section{Numerical results}
\label{sec-numeric}

\begin{figure}
\includegraphics[width=8.6 cm]{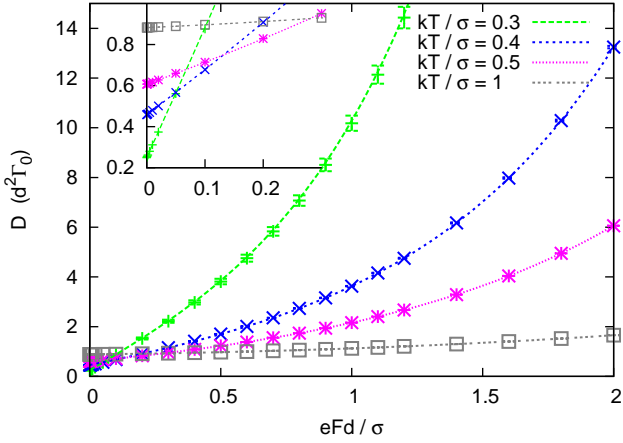}
\caption{$D(F)$ in the RBM for different temperatures $T$. The curves
show the analytical solution Eq.~(\ref{eq:2-D-RBM}), while the symbols show
numerical results (Eq.~\ref{eq-Derrida-D}) for chains with $N=10^7$ sites.
The inset shows the low-field behavior.
}
\label{fig-1d-RB}
\end{figure}

\begin{figure}
\includegraphics[width=8.6 cm]{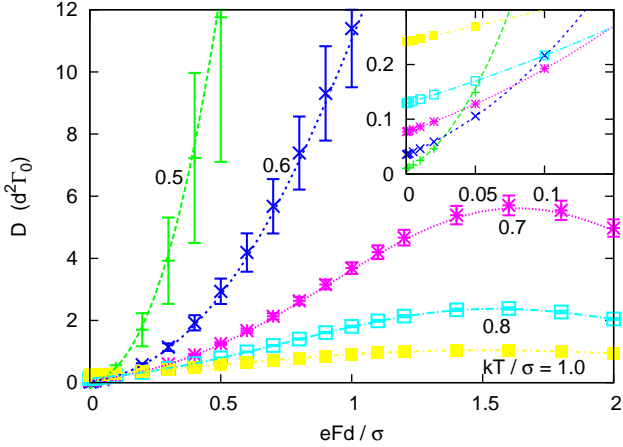}
\caption{$D(F)$ in the REM for different temperatures $T$.
The curves show the analytical solution (Eq.~(\ref{eq:2-D-from-m-REM}) with
$m_1 \dots  m_{11}$ from the MA column in Table~\ref{tab:2-m}).
The symbols show numerical results (Eq.~\ref{eq-Derrida-D})
for chains with $N=10^8$ sites.
The inset shows the low-field behavior.
}
\label{fig-1d-RE}
\end{figure}

In order to verify the analytical results obtained above, we
perform numerical calculations for a one-dimensional chain of $N$
hopping sites using the equations that give the drift velocity $v$
and the diffusion coefficient $D$ as a function of all the hopping
rates $\Gamma_{i \pm 1, i}$ in the chain \cite{Derrida1983}:
\begin{align}
\label{eq-Derrida-D}
D = {} &  \frac {1}{(\sum_{n=1}^N r_n)^2}
\Bigg(
v \sum_{n=1}^N u_n \sum_{i=1}^N ir_{n+i}
 + \notag\\
& N \sum_{n=1}^N \Gamma_{n, n+1}\ u_n r_n
\Bigg)
\ - \ v \frac{N+2}{2},
\end{align}
\begin{equation}
v =  \frac {N}{\sum_{n=1}^N r_n}
\left[1 - \prod_{n=1}^N \left(\frac{\Gamma_{n+1,n}}{\Gamma_{n,n+1}} \right)
\right],
\end{equation}
\begin{equation}
r_n = \frac{1}{\Gamma_{n,n+1}} \left[
1 + \sum_{i=1}^{N-1} \prod_{j=1}^{i}
\left( \frac{\Gamma_{n+j,n+j-1}}{\Gamma_{n+j,n+j+1}} \right)
\right],
\end{equation}
\begin{equation}
u_n = \frac{1}{\Gamma_{n,n+1}} \left[
1 + \sum_{i=1}^{N-1} \prod_{j=1}^{i}
\left( \frac{\Gamma_{n-j+1,n-j}}{\Gamma_{n-j,n-j+1}} \right)
\right].
\label{eq-Derrida-u}
\end{equation}
In this section, we follow Ref.~\onlinecite{Derrida1983} and set
the distance $d$ between sites equal to unity for the sake of
simplicity.

 For both the RBM and the REM, the diffusion
coefficient was obtained for different temperatures and fields, by
generating several chains with random jump rates (according to the
respective model), evaluating Eq.~(\ref{eq-Derrida-D}) for each
chain and averaging the results.  Long chains ($10^7$ and $10^8$
sites) were needed to obtain a good agreement between different
realizations of the chains.

For chains of this length, a direct evaluation of
Eqs.~(\ref{eq-Derrida-D})--(\ref{eq-Derrida-u}) is not practical.
Below, the equations are rewritten in a form that can be evaluated
in $O(N)$ steps, using recursion relations. Define
\begin{equation}
g_n = \frac{\Gamma_{n,n-1}}{\Gamma_{n,n+1}}
 \quad \text{ and } \quad
h_n = \frac{\Gamma_{n+1,n}}{\Gamma_{n,n+1}},
\end{equation}
 and further
\begin{align}
G_n &= 1 + \sum_{i=1}^{N-1} \prod_{j=1}^{i} g_{n+j} \\
H_n &= 1 + \sum_{i=1}^{N-1} \prod_{j=1}^{i} h_{n-j},
\end{align}
so that $r_n = G_n \ / \ \Gamma_{n,n+1}$ and
$u_n = H_n \ / \ \Gamma_{n,n+1}$.
All $G_n$ and $H_n$, and thus $r_n$ and $u_n$ can now be calculated
efficiently from
\begin{equation}
\label{eq-recursion-G}
G_{n-1} =  g_n G_n - G + 1,
\end{equation}
\begin{equation}
H_{n+1}=h_n H_n - H + 1
\end{equation}
where $G = H = g_1 g_2 \ldots g_N = h_1 h_2 \ldots h_N$.
For the first term in the brackets in Eq.~(\ref{eq-Derrida-D}), define
$S_n = \sum_{i=1}^N i r_{n+i}$ and $S = \sum_{i=1}^N r_{n}$.  Now
\begin{equation}
S_{n+1} = S_n - S + N r_{n+1}.
\end{equation}
These relations are numerically
stable if $G < 1$,
which is satisfied if the average drift is to the right (towards
larger site indices).
The diffusion coefficient is now given by
\begin{equation}
\label{eq-D-Derrida-fast}
D = \frac{1}{S^2} \left(
v \sum^N_{n=1} \frac{H_n S_n}{\Gamma_{n,n+1}} +
N \sum^N_{n=1} \frac{G_n H_n}{\Gamma_{n,n+1}}
\right) - v \frac{N+2}{2}.
\end{equation}
It seems tempting to write Eq.~(\ref{eq-recursion-G}) in the form
\[
G_n = (G_{n-1} + G - 1)/ g_n,
\]
so that all equations could be evaluated starting from $n = 1$,
but this form is too susceptible to numerical errors to be usable in
practice.  Thus one has to evaluate all $G_n$ with Eq.~(\ref{eq-recursion-G})
starting from $G_N$ and
store them in a table. $S_n$ and $H_n$ do not need to be stored, since 
they can be evaluated
while performing the sum in Eq.~(\ref{eq-D-Derrida-fast}),
starting from $n=1$.

With this method of evaluation,
numerical results for the diffusion coefficient were obtained.
The results are shown in
Fig.~\ref{fig-1d-RB} for the RBM, together with the analytical results,
and in Fig.~\ref{fig-1d-RE} for the REM.
For both models the diffusion coefficient is linear in the electric
field (at low fields), see the insets in each figure.

\section {Discussion and conclusions}
\label{sec-conclusions}

Figures~\ref{fig-1d-RB} and~\ref{fig-1d-RE} demonstrate a full
agreement between analytical results, Eqs.~(\ref{eq:2-D-RBM})
and (\ref{eq:2-D-from-m-REM}), and numerical ones based on
Eq.~(\ref{eq-Derrida-D}). This result can be considered as
evidence that the two definitions of the diffusion coefficient $D$,
Eq.~(\ref{eq:2-def-D1}) and Eq.~(\ref{eq:2-def-D2}), give the same
quantity for hopping in one-dimensional disordered systems. The
former definition expresses $D$ via the variance of particle
displacement during a random walk, while the latter one defines
$D$ as a ratio between the particle flow and the gradient of
macroscopic concentration of particles. Although it seems to be
obvious from a physical point of view that both definitions should
give the same result, no formal proof has yet been known.

Computer simulations evidence that at low temperatures the
diffusion constant experiences significant fluctuations from one
realization to another, even for systems containing millions of
localized states. The reason of such a large fluctuations in 1D
hopping is essentially the same as in 3D hopping---it is the
sensitivity of the diffusion coefficient to the very rare sites
with low energies. We will discuss this phenomenon in detail in
the following paper.\cite{Nenashev2009II}
We also note that sample-to-sample fluctuations of \emph{mobility}
for all parameters presented in Figures~\ref{fig-1d-RB}
and~\ref{fig-1d-RE} are negligible.

The most striking property of the 1D diffusion is its linear
dependence on electric field:
\begin{equation} \label{eq-discussion-linear}
 D(F) = D(0) + \alpha |F| + O(F^2) \,,
\end{equation}
where $\alpha\neq0$. It means that the diffusion coefficient is a
\emph{non-analytic} function of electric field. From general
physical arguments one can hardly expect such a behavior. Instead,
one can expect that, as $D(F)$ is an even function, it can be
expanded in a Taylor series with respect to $F^2$:
\begin{equation} \label{eq-discussion-parabolic}
 D(F) = D(0) + \beta F^2 + O(F^4) \,.
\end{equation}
The physical reason of this non-analyticity is still unclear. As
first steps to acquire an understanding of this phenomenon, we will
try  (i)~to provide a \emph{mathematical} explanation of it and
(ii)~to find out which systems demonstrate the linear field
dependence of the diffusion coefficient and which systems lack
this behavior.

From the mathematical point of view, the possibility for a
non-analytic dependence $D(F)$ can be seen from expressions
(\ref{eq:2-solution-for-a-energy}) and
(\ref{eq:2-solution-for-b-energy}) for the coefficients $a_i$ and
$b_i$ as functions of $B\equiv\exp(eFd/kT)$. These expressions are
series that converge at $B>1$ and diverge at $B\leq1$. For $B<1$
(i.~e.~for negative $eF$) one can obtain converging series,
applying a ``mirror reflection'' transformation ($i+k\rightarrow
i-k$, $B\rightarrow B^{-1}$) to
Eqs.~(\ref{eq:2-solution-for-a-energy}) and
(\ref{eq:2-solution-for-b-energy}). Therefore $a_i$ and $b_i$ are
defined by \emph{different} series expansions for positive and
negative values of the field. Moreover, at $eF>0$ the values $a_i$
and $b_i$ depend on quantities related to sites $i$, $i+1$, $i+2$,
... (for example, on $\Gamma_{i+1,i+2}$, $\varepsilon_{i+1}$,
etc.); on the contrary, for $eF<0$ these coefficients $a_i$ and
$b_i$ depend on a different set of sites: $i$, $i-1$, $i-2$, ...
So, it is obvious that, in a disordered system, the function
$a_i(F)$ for negative $F$ cannot be obtained by analytic
continuation of this function for positive $F$, and vice versa.
The same is true for $b_i(F)$. Keeping in mind that the mobility
$\mu\equiv v/F$ and the diffusion coefficient $D$ depend on $a$'s and
$b$'s via Eq.~(\ref{eq:2-vD-from-ab}), one can conclude that
negative-field parts of the functions $\mu(F)$, $D(F)$ may not be
analytic continuations of their positive-field parts.
Consequently, a non-analyticity of $\mu(F)$ and $D(F)$ at $F=0$ is
possible.

It is evident from Figs.~\ref{fig-1d-RB} and \ref{fig-1d-RE} and from
Eqs.~(\ref{eq:2-D-RBM}) and (\ref{eq:2-approx_D}) that both the RBM
and the REM demonstrate a linear low-field behavior of $D(F)$, according
to Eq.~(\ref{eq-discussion-linear}). Moreover,
Eq.~(\ref{eq:2-approx_mu}) shows that the mobility $\mu(F)$ in the REM
also contains a linear contribution with respect to $F$, while in the
RBM the mobility is a smooth function of $F$, see
Eq.~(\ref{eq:2-v-RBM}).

Parris \emph{et al.}\cite{Parris1997,Parris1997a} considered a
model of a 1D continuous medium with smooth disorder potential, and
obtained analytical expressions for $\mu(F)$ and $D(F)$. Although
the low-field behavior of the mobility and of the diffusion
coefficient were not discussed in
detail,\cite{Parris1997,Parris1997a} one can learn from equations
(25) and (69) of Ref.~\onlinecite{Parris1997a} that this behavior
is qualitatively the same as in our REM. The method of 
Refs.~\onlinecite{Parris1997,Parris1997a} is, however, not
directly applicable to the Gaussian disorder model considered here.
Therefore a separate derivation was necessary.

One-dimensional transport with Gaussian density of states (DOS)
has another peculiarity---namely,  there are sites with
arbitrarily high energies, which represent barriers for
transport. Despite the fact that such barriers are very rare, their
influence on the transport properties should be noticeable (unlike in the 2D
and 3D cases, where a carrier can easily pass around these hard
places). One can argue that this peculiarity might be responsible
for the unusual non-analytical behavior of $\mu(F)$ and $D(F)$. In order to
check this assumption, we consider a 1D system with a \emph{discrete}
DOS $g(\varepsilon)$ allowing only two values of energy
($\varepsilon=-\sigma$ and $\varepsilon=\sigma$):
\begin{equation}
 g(\varepsilon) = \frac12 \delta(\varepsilon+\sigma) + 
\frac12 \delta(\varepsilon-\sigma).
\end{equation}
It is obvious that there are no high barriers in this system.  Using
our general expressions 
(\ref{eq:2-v-from-m-RBM}), (\ref{eq:2-D-from-m-RBM}),
(\ref{eq:2-v-from-m-REM}), and (\ref{eq:2-D-from-m-REM}),
we have calculated the dependencies $\mu(F)$ and $D(F)$ for this DOS
for both RBM and REM. The results demonstrate qualitatively the same
low-field behavior as in the case of Gaussian DOS. Therefore the
property of the non-analyticity of $\mu$ and $D$ cannot be attributed
just to the presence of infinitely high barriers provided by the
corresponding DOS.

We have also checked whether this non-analyticity is related to the
assumption of the \emph{nearest-neighbor} hopping.  We have considered
the same random-energy model as discussed in
Sec.~\ref{sec-analytic-REM}, except that we allow transitions to
distant states.  The dependence of transition rates $\Gamma_{ij}$ on
distances $r_{ij}$ between the sites is governed by the Miller-Abrahams
expression~(\ref{eq-Miller-Abrahams}).  The mobility and diffusion
coefficient as functions of electric field were calculated by a
Monte-Carlo algorithm described in the following paper.\cite{Nenashev2009II}
Again, the results have
shown linear low-field dependencies $\mu(F)$ and $D(F)$.

Therefore one can conclude that the linear behavior of the diffusion
coefficient, Eq.~(\ref{eq-discussion-linear}), is a robust property of
1D disordered systems.

On the contrary, at higher dimensions there are no evidences of a linear
field dependence of the diffusion coefficient. Both our Monte-Carlo
simulations (see the following paper, Ref. \onlinecite{Nenashev2009II}) 
and previous studies\cite{Pautmeier1991} clearly
demonstrate a parabolic field dependence,
Eq.~(\ref{eq-discussion-parabolic}), for 2D and 3D systems with
Gaussian disorder. It is worth to note that a smooth dependence $D(F)$
should be obtained also in 1D systems \emph{without
  disorder}---namely, in systems with periodically repeated site
energies (and barrier heights). Indeed, such a system is equivalent to
a \emph{finite} chain (with periodical boundary conditions) considered
by Derrida.\cite{Derrida1983} One can therefore use Derrida's
formula for $D$, Eq.~(\ref{eq-Derrida-D}). This formula, in the case
of finite chain length $N$, is an analytic function of the transition
rates $\Gamma_{ij}$, and consequently of the electric field. Our numerical
studies also confirm that for small $N$ there is a region of parabolic
dependence $D(F)$ around $F=0$. Thus, disorder is important for the 
non-analytic behavior of $D(F)$.

Finally, we will discuss the applicability of Einstein's relation
(\ref{eq-Einstein}) for finite electric fields.  Our analytical results,
Eqs.~(\ref{eq:2-v-RBM}), (\ref{eq:2-D-RBM}), (\ref{eq:2-approx_mu}),
and (\ref{eq:2-approx_D})
show that Einstein's relation is violated at any
non-zero field $F$ both in RBM and REM, and that the deviation from
Eq.~(\ref{eq-Einstein}) is proportional to $|F|$. One should note that
this phenomenon is related to the discrete nature of the systems, in which
charge transport is dominated by hopping processes. For the opposite
case of a continuous-medium model, the diffusion coefficient is known
to obey a generalized version of the Einstein
relation:\cite{Parris1997,Parris1997a}
\begin{equation} \label{eq-discussion-gen_Einstein}
 D(F) = \frac{kT}{e} \, \frac{dv}{dF} \,,
\end{equation}
where $v(F)$ is the drift velocity. Our expressions for both RBM and
REM, however, demonstrate that Eq.~(\ref{eq-discussion-gen_Einstein})
is also violated in the general case, and the deviation is also
proportional to $|F|$. We argue that, generally, there are no exact
connections between $D(F)$ and $\mu(F)$ at $F\neq0$. Indeed, let us
consider the REM.  There are eleven quantities $m_1\ldots m_{11}$ (see
Sec.~\ref{sec-analytic-REM}) dependent on statistics of the energy levels
and on the electric field.  The mobility depends on three of them
($m_1,m_3,m_4$), according to Eq.~(\ref{eq:2-v-from-m-REM}); the diffusion
coefficient depends on all eleven quantities, see
Eq.~(\ref{eq:2-D-from-m-REM}). For an arbitrary density of states, all
the eleven quantities are independent of each other and cannot be
reduced to each other. Consequently, there is no general way to
reduce the diffusion coefficient to the mobility at non-zero electric field.

In conclusion, we have examined analytically and numerically two
models of one-dimensional hopping  transport---the random-barrier
(RBM) and the random-energy model (REM). Exact analytical solutions
of field-dependent diffusion coefficient have been obtained for
both models in the case of nearest-neighbor hopping. We have
demonstrated that the non-analytic field dependence
(\ref{eq-discussion-linear}) of the diffusion coefficient, as well as
the violation of the Einstein relation for any nonzero electric
field, are inherent properties of hopping transport in
one-dimensional disordered systems.

\begin{acknowledgments}
We are indebted to Prof. Boris Shklovskii for valuable
discussions. Financial support from the Academy of Finland project
116995, from the Deutsche Forschungsgemeinschaft and that of the
Fonds der Che\-mischen Industrie is gratefully acknowledged. Parts
of the calculations were done at the facilities of the Finnish IT
center for science, CSC.
\end{acknowledgments}

\newpage
%\bibliography{diffusion}

\end{document}